\documentclass[sigconf,nonacm]{acmart}

\AtBeginDocument{%
  }

\setcopyright{none}
\usepackage[noend]{algpseudocode}
\usepackage{algorithm}
\usepackage{bbm}
\usepackage{graphicx}
\usepackage[labelformat=simple]{subcaption}
\usepackage{listings}
\usepackage[compress]{cleveref}
\usepackage{mathrsfs}
\usepackage{bm}
\usepackage{balance}
\usepackage{outlines}
\usepackage{graphics}
\usepackage{svg}
\usepackage{grffile}
\usepackage{xspace}
\usepackage{enumitem}
\usepackage{xcolor}
\usepackage{array}
\usepackage{nccmath}
\usepackage{multirow}
\usepackage{pifont}
\usepackage{soul}
\usepackage{caption}
\usepackage{upgreek}

\usepackage{xpatch}
\xpatchcmd{\NCC@ignorepar}{%
\abovedisplayskip\abovedisplayshortskip}
{%
\abovedisplayskip\abovedisplayshortskip%
\belowdisplayskip\belowdisplayshortskip}
{}{}

\graphicspath{./figures/}

\crefformat{section}{\S#2#1#3} 
\crefformat{subsection}{\S#2#1#3}
\crefformat{subsubsection}{\S#2#1#3}
\crefname{section}{Section}{Sections}
\Crefname{section}{Section}{Sections}
\crefname{figure}{Figure}{Figures}
\Crefname{figure}{Figure}{Figures}
\crefname{subfigure}{Figure}{Figures}
\Crefname{subfigure}{Figure}{Figures}
\crefname{algorithm}{Algorithm}{Algorithms}
\Crefname{algorithm}{Algorithm}{Algorithms}
\crefname{table}{Table}{Tables}
\Crefname{table}{Table}{Tables}
\crefrangelabelformat{subfigure}{#3#1#4--#5(\crefstripprefix{#1}{#2}#6}

\newcommand{\mi}[1]{[\textcolor{orange}{[MI:] #1}]}

\definecolor{emerald}{rgb}{0.31, 0.78, 0.47}

\newcommand{\eat}[1]{}

\newcommand{\stitle}[1]{\vspace{0.1ex}\noindent{\bf #1}}

\newcommand{\preeq}{\vspace{0mm}\begin{small}}
\newcommand{\posteq}{\vspace{0mm}\end{small}}

\newcommand{\system}{\textsc{TDP}\xspace}
\newcommand{\lsystem}{\textsc{TQP}\xspace}

\newcommand{\mnistgrid}{MNISTGrid\xspace}

\usepackage{url}

\DeclareFixedFont{\ttb}{T1}{txtt}{bx}{n}{8}
\DeclareFixedFont{\ttm}{T1}{txtt}{m}{n}{8}

\usepackage{color}
\definecolor{deepblue}{rgb}{0,0,0.5}
\definecolor{deepred}{rgb}{0.6,0,0}
\definecolor{deepgreen}{rgb}{0,0.5,0}
\definecolor{purple}{rgb}{0.5,0,0.5}
\definecolor{gray}{rgb}{0.33,0.33,0.33}

\definecolor{dkgreen}{rgb}{0,0.6,0}
\definecolor{gray}{rgb}{0.5,0.5,0.5}
\definecolor{mauve}{rgb}{0.58,0,0.82}

\lstdefinelanguage{Python}{
	keywords={typeof, torch, nonzero, index_select, zeros_like, lt, masked_select, new, true, false, catch,def,val, function, return, null, catch, switch, var, shape,  while, do, else, case, break, override, range},
	keywordstyle=\color{blue}\bfseries,
	ndkeywords={class, export,extends, boolean, throw, implements, import, this, abstract, for, in, if, from, select, avg, where},
	ndkeywordstyle=\color{dkgreen}\bfseries,
	otherkeywords={+, =>,<=, ==, >,< , ||},
	identifierstyle=\color{black},
	sensitive=false,
	comment=[l]{\#},
	morecomment=[s]{"""}{"""},
	commentstyle=\color{purple}\ttfamily,
	stringstyle=\color{red}\ttfamily,
	morestring=[b]',
	morestring=[b]"
}
\begin{document}

\title{
The Tensor Data Platform: Towards an AI-centric Database System}

\author{Apurva Gandhi, Yuki Asada, Victor Fu, Advitya Gemawat, Lihao Zhang, Rathijit Sen\\Carlo Curino, Jes\'us Camacho-Rodr\'iguez, Matteo Interlandi}

\affiliation{
    \institution{Microsoft}
    \country{}
}
\email{{<firstname>.<lastname>}@microsoft.com}

\begin{abstract}



Database engines have historically absorbed many of the innovations in data processing, adding features to process graph data, XML, object oriented, and text among many others. In this paper, we make the case that it is time to do the same for AI---but with a twist! While existing approaches have tried to achieve this by integrating databases with external ML tools, in this paper we claim that achieving a truly AI-centric database requires moving the DBMS engine, at its core, from a relational to a tensor abstraction. This allows us to: (1) support multi-modal data processing such as images, videos, audio, text as well as relational; (2) leverage the wellspring of innovation in HW and runtimes for tensor computation; and (3) exploit automatic differentiation to enable a novel class of~``trainable'' queries that can learn to perform a task. 

To support the above scenarios, we introduce \system: a system that builds 
upon our prior work mapping relational queries to tensors. 
Thanks to a tighter integration with the tensor runtime, \system is able to provide a broader coverage of 
new emerging scenarios requiring access to multi-modal data and automatic differentiation.

\end{abstract}

\maketitle

\begin{sloppypar}

\section{Introduction}

Relational database engines have dominated the data processing landscape for almost 50~years, integrating many of the new ideas in data processing as ``features'' of the existing core engine. Though recently, new Machine Learning (ML) systems have emerged to support processing data that is not as naturally mapped to the relational model (e.g., video, images, audio, large text, high-dimensional vector data).
Many of these systems support neural network training and inference and are underpinned by Tensor Computation Runtimes (TCRs) such as PyTorch \cite{pytorch} and TensorFlow \cite{tensorflow}. Investments in these runtimes and specialized HW to accelerate them is tracking the insatiable market hunger for ``AI'' tech. Venture capitalists alone are pouring \$2B/quarter in special-HW for neural networks~\cite{ai-hw-market}. 
Interestingly, an ever-growing number of organizations are embracing \textit{mixed workloads} that combine multiple such systems into one workflow to satisfy the requirements of a large class of emerging applications~\cite{raven-sigmod,analyticdb-v,9705193}. Specifically prominent are scenarios that combine relational processing and ML. 


The recent heavy investments in ML have led to a thriving ecosystem of open source TCRs 
that are leveraged by data scientists and software developers to implement and run their models efficiently. These libraries have certain characteristics that make them appealing as the target for a wide variety of workloads, namely: (1)~they use specialized kernels that run efficiently on CPU, 
GPU, 
but are also capable of leveraging the latest hardware accelerators such as TPU, 
Cerebras, 
and IPU; 
(2)~their data model, based on the \textit{tensor} abstraction~\cite{hummingbird-vision}, is flexible enough to represent multiple data modalities through embeddings, including tables, text, graphs, images, or videos; (3)~they have a rich and composable API 
providing a declarative interface enabling complex computations (while hiding low-level implementation details), as well as novel features such as \emph{automatic differentiation}.

Integrating relational and ML workloads has been studied since the early '90s~\cite{sqlserver-data-mining}, leading to numerous works (e.g.,~\cite{madlib,raven,raven-sigmod,masq,10.14778/3317315.3317323,10.1145/3468791.3468840}) 
that have proposed different techniques over the years, ranging from integrating ML as a UDF through an external specialized system, to expressing ML algorithms directly in SQL. Most of the proposals follow a common theme: ML is merely a guest in the relational house owned by the DBMS. This has two fundamental limitations: (1) it is poorly suited to handle non-relational data, and (2) it misses out on the virtuous cycle among HW vendors/OSS/ML academics/app developers that TCR engines enjoy.

In this paper, we argue that an alternative path exists where we embrace the technologies developed by (and for) the ML community from the ground up, and put them at the core of the database runtime to unlock new capabilities and synergies. We show that the resulting systems can handle: (1)~legacy relational workloads, (2)~specialized use-cases such as graphs and ML, and (3)~emerging applications such as vector search over images and audio, and video analytics. Our implementation of one such system, which we refer to as the \emph{Tensor Data Platform} (\system)~(\S\ref{sec:in-ml-db}), leverages PyTorch to run queries over structured and unstructured data on a wide range of hardware devices. \system integrates the flexibility of PyTorch's programming model with the declarative power of SQL~(\S\ref{sec:udf}), leading to a hybrid ML-SQL experience that is appealing to database users without forcing data scientists outside of their comfort zone (e.g., Python). Importantly, the tight integration with PyTorch allows \system to support \textit{trainable queries}~(\S\ref{sec:declarative_ml}) that leverage automatic differentiation built into the system~\cite{pytorch} to train models embedded in them. Overall, we demonstrate that \system facilitates the implementation and efficient execution of a wide variety of applications, in different domains, using a single unified system~(\S\ref{sec:use-cases}).

The Tensor Data Platform is our answer to the observation that several untapped possibilities lie in the intersection between ML and database systems. We hope that the community will join us in the journey of redesigning databases towards an AI-centric system.

\section{\system: An AI-centric Database System}
\label{sec:in-ml-db}

\system is a data processing platform implemented on top of the tensor data structure and TCRs like PyTorch. \system is completely written in Python, and includes an integrated query processor (similar to DuckDB~\cite{duckdb}) leveraging PyTorch for hardware acceleration, automatic differentiation, and support for unstructured data. 
\system's query processor extends \lsystem~\cite{tqp,hummingbird}, a system that we introduced previously, to fully leverage PyTorch's capabilities beyond the execution of relational queries it originally supported.
Finally, \system naturally blends with the ML ecosystem and tools such as Notebooks, TensorBoard, Pandas, Numpy, etc.~\cite{tqp-demo}. In the following, we describe three key features of \system: a generic \emph{storage model} for structured and unstructured data; support for \emph{data encoding} schemes allowing to seamlessly move across different data modalities; and a flexible \emph{query processor}.

\eat{
\system is a query processor implemented on top of PyTorch. Concretely, \system extends \lsystem~\cite{tqp,hummingbird}, a system that we introduced previously, to fully leverage PyTorch's capabilities beyond the execution of relational queries it originally supported. 
\system is completely written in Python, it is an integrated query processor similar to DuckDB~\cite{duckdb}, but by design it leverages PyTorch for hardware acceleration, automatic differentiation, and support for unstructured data. In addition, \system naturally blends with the ML ecosystem and tools such as Notebooks, TensorBoard, Pandas, Numpy, etc.~\cite{tqp-demo}.
In the following, we describe four key features of \system: generic \emph{storage model} for structured and unstructured data, support for \emph{data encoding}, integration with external systems for \emph{query compilation}, and flexible \emph{query execution}.}

\stitle{Storage Model.} 
\system stores relational data in a columnar format, where each column is a PyTorch \emph{tensor}. Tensors are multidimensional arrays with an arbitrary numbers of dimensions. As such, \system can store tabular data as a collection of 1-d tensors (i.e., each column is viewed as a vector), but it also supports columns containing 2-d tensors (i.e., each row containing a vector), 3-d tensors (e.g., each row containing a gray scale image), 4-d tensors (e.g., each row contains an rgb image), etc.
Thanks to this design, \system can natively store both structured and unstructured data, and importantly, it can provide a unified view of data such that mixed scalar-vector queries~\cite{milvus,analyticdb-v} can be both expressed in a natural way and executed efficiently. 
\system accepts input data in different formats.
When data is registered into \system, 
it is first transformed into tensors and subsequently \emph{encoded}. Data can be stored both on CPU and GPU.
\system focuses on analytical workloads, whereby currently there is no transactional support.

\stitle{Data Encoding.}
One of the key features of columnar databases is the ability to encode data into compressed formats that can both decrease memory requirements and increase query performance. 
Similarly, \system does not use PyTorch tensors directly, but rather provides its own \emph{encoded tensors} abstraction, i.e., tensors with attached metadata describing how data is stored in them. 
\system for the moment uses \emph{plain encoding} for numerical data, \emph{order-preserving dictionary encoding} for string columns (where the dictionary itself is a 2-dimensional plain tensor, storing one string-vector per row), and \emph{Probability Encoding}~(PE) which attaches structured information to numerical data (more on this in the next sections). 
Similar to columnar databases, \system leverages the metadata information of encoded tensors to pick the right execution strategy for operators. 
\system provides an \texttt{encode}/\texttt{decode} APIs to easily move back and forth between the encoded and decoded formats.

\begin{example}[Ingesting Data]
We start by loading some simple tabular data in \system. The data is in a Pandas dataframe, and therefore we can use the \texttt{register\_df} API to store it in \system. Under the hood, \system takes care of converting, encoding, and moving the data to the requested device. 
Similar APIs exist for registering multidimensional NumPy arrays, Arrow arrays, Parquet files and, of course, PyTorch tensors.
These APIs are generic enough for supporting both structured and unstructured data.

\lstloadlanguages{Python}
\lstset{frame=tb,
	language=Python,
	commentstyle=\color{dkgreen},
	aboveskip=3mm,
	belowskip=3mm,
	showstringspaces=false,
	columns=flexible,
	basicstyle={\footnotesize\ttfamily},
	numberstyle=\tiny\color{gray},
	keywordstyle=\color{blue},
	stringstyle=\color{mauve},
	breaklines=true,
	breakatwhitespace=true,
	tabsize=3,
	numbers=none,
	xleftmargin=1em,
	framexleftmargin=1em,
}

\begin{lstlisting}[aboveskip=\smallskipamount,belowskip=\smallskipamount,caption=Data ingestion in \system: a Pandas dataframe \texttt{data} is stored as a \texttt{numbers} table in GPU memory., label={lst:register_data}]
tdp.sql.register_df(data,"numbers", device="cuda")
\end{lstlisting}
\end{example}


\stitle{Query Processor.}
\system leverages external query parsers and optimizers for generating physical plans. Currently, \system can rely on Spark~\cite{spark-sql} and Substrait~\cite{substrait} for this purpose.   
Once the physical plan is generated, \system compiles it into a sequence of PyTorch models, one per operator in the physical plan. 
\system contains an internal dictionary of PyTorch models, each of them implemented using PyTorch's tensor API. For each physical operator, we can have more than one PyTorch implementation, and at compilation time we use a mix of flags (e.g., Listing~\ref{lst:trainable_query}) and heuristics to pick which one to use. More details on the compilation phase, supported operators, and how to express relational operators using the PyTorch's tensor API can be found in~\cite{tqp}.

\begin{example}[Query Compilation]\label{ex:query-compilation}
We submit an aggregate query (line~1 in Listing~\ref{lst:compile_query}) over the previously registered {\sf numbers} table. 

\begin{lstlisting}[aboveskip=\smallskipamount,belowskip=\smallskipamount,caption=Query definition and compilation in \system., label={lst:compile_query}]
statement = "SELECT Digits, Sizes, COUNT(*)                             FROM numbers GROUP BY Digits, Sizes"
compiled_query = tdp.sql.spark.query(statement, device="cuda")
\end{lstlisting}
\end{example}


The output of query compilation is a PyTorch model and, as such, it can be for example: 
used in a training loop (more on this in \S\ref{sec:udf}), executed on different hardware devices (in the example we compiled the query for GPU execution), further optimized using compilers such as TVM, profiled using Tensorboard~\cite{tqp-demo}, etc.

\begin{example}[Query Execution]
Now that we have compiled the query, we can execute it as shown in Listing~\ref{lst:execute_query}, where we ask \system to generate the output in Pandas dataframe format.
\system execution API is flexible enough to support data modalities beyond tabular. For instance, we can also generate 
outputs which can be rendered into images using Matplotlib, or audio using {\sf IPython.display.Audio}.

\lstloadlanguages{Python}
\lstset{frame=tb,
	language=Python,
	commentstyle=\color{dkgreen},
	aboveskip=3mm,
	belowskip=3mm,
	showstringspaces=false,
	columns=flexible,
	basicstyle={\footnotesize\ttfamily},
	numberstyle=\tiny\color{gray},
	keywordstyle=\color{blue},
	stringstyle=\color{mauve},
	breaklines=true,
	breakatwhitespace=true,
	tabsize=3,
	numbers=none,
	xleftmargin=1em,
	framexleftmargin=1em,
}

\begin{lstlisting}[aboveskip=\smallskipamount,belowskip=\smallskipamount,caption=Executing the compiled query in GPU and returning the result into a Pandas dataframe., label={lst:execute_query}]
result = compiled_query.run(toPandas=True)
\end{lstlisting}
\end{example}

\begin{figure*}[t!]
\begin{tabular}{p{0.65\textwidth}p{0.35\textwidth}}
    \hspace{-3ex}
    \begin{minipage}{.65\textwidth}
    \centering
    \includegraphics[width=\linewidth]{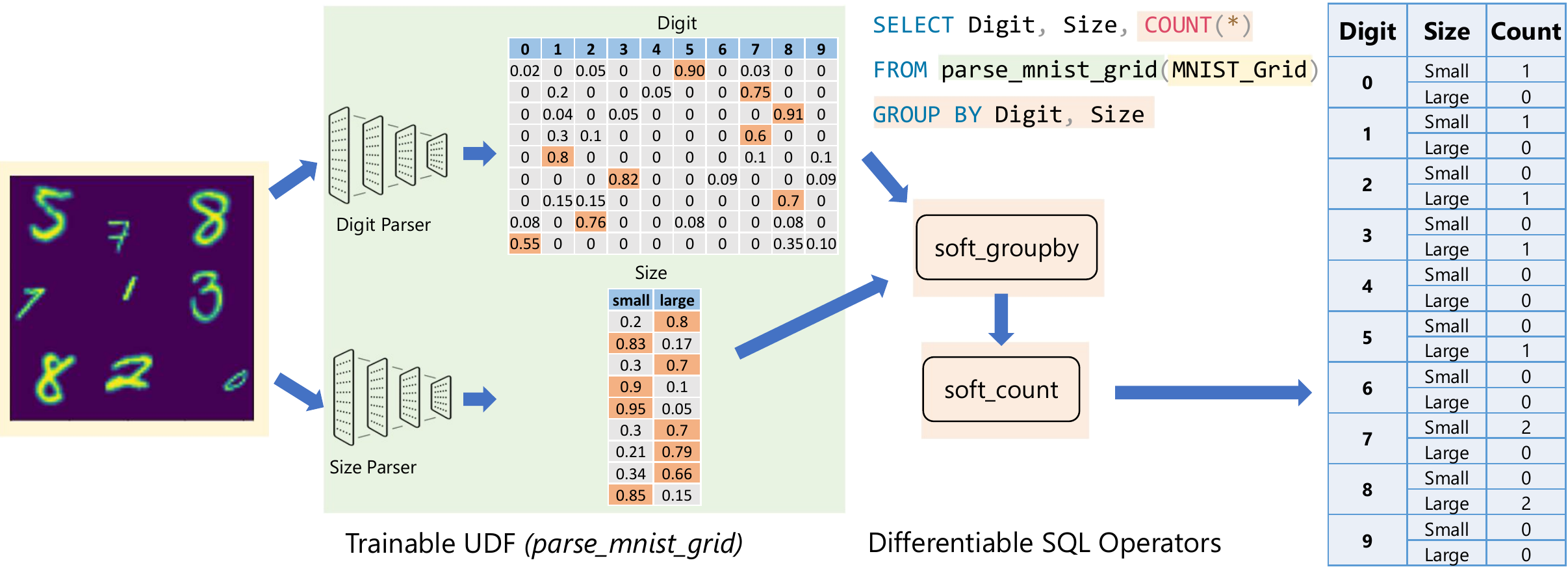}
    \caption{Anatomy of the query execution plan trained on the MNISTGrid dataset. In yellow: an \mnistgrid image. In green: the trainable TVF generating  probability vectors from the images. In orange: ``soft'' operators implementing the {\sc group by} and {\sc count} logic. On the right side we show the desired output of the query.}
    \label{fig:mnist-grid}
    \end{minipage}
    &
    \hspace{-4.5ex}
    \begin{minipage}{.35\textwidth}
        \vspace{-2ex}
    \begin{lstlisting}[caption=TVF to parse MNISTGrid into the structured format supported by \system's SQL., label={lst:mnistgrid}]
digit_parser = CNN(num_classes=10)
size_parser = CNN(num_classes=2)

@tdp_udf("Digit float, Size float")
def parse_mnist_grid(mnist_grid: torch.Tensor) -> Tuple[torch.Tensor]:
    # Break up grid into a batch of 9 tiles/images
    tiles = einops.rearrange(
        mnist_grid,
        "1 (h1 h2) (w1 w2) -> (h1 w1) 1 h2 w2", h1=3, w1=3
    )
    # return digit and size classification outputs
    return PEEncoding.encode(digit_parser(tiles)), PEEncoding.encode(size_parser(tiles))
\end{lstlisting}
    \end{minipage}
    \end{tabular}
    \vspace{-0ex}
\end{figure*}

\vspace{-1.5ex}
\section{ML-first User Experience}
\label{sec:udf}
Since \system lowers its SQL query execution plan to PyTorch, a powerful implication is that SQL execution 
has complete interoperability with 
PyTorch. 
In this section, we provide three design choices we embraced for surfacing familiar patterns to ML practitioners, from within the database.

\stitle{\emph{ML within SQL}: UDF-based programming model.}
In \system users can surface custom PyTorch code within a SQL query through User-Defined and Table-Valued Functions (UDFs/TVFs). These functions can encapsulate arbitrarily complex ML models, e.g., to parse unstructured data into a structured representation on which SQL operators can be applied. 
Data is passed into UDFs/TVFs as (encoded) tensors, and \system expects (encoded) tensors as results.
\system provides an annotation API simplifying the process of registering Python functions into the framework. 
While UDFs and TVFs have already been explored to add ML features to SQL systems (e.g.,~\cite{madlib}), the novelty of our approach is that we do not use them for calling into external tools, but rather as a means to access the underlying TCR API. 
In the end,  
UDFs/TVFs and SQL operators are all compiled down into PyTorch programs.

\begin{example}[\mnistgrid]

We want to extend the SQL query in Example~\ref{ex:query-compilation} to work over a grid of handwritten digit images rather than a table. 
Throughout this example, we use a variant of the MNIST handwriting digit dataset, 
which we refer to as \mnistgrid, containing 9x9 grids of (small/large) resized handwritten digits.
Fig.~\ref{fig:mnist-grid} summarizes the workflow of our approach.
%
\system allows to achieve our goal with little modification to the original query. We simply call a \texttt{parse\_mnist\_grid} TVF, shown in Listing~\ref{lst:mnistgrid}, 
to parse \mnistgrid images to a structured format. Functions are registered in \system using the \texttt{tdp\_udf} annotation (line~4). 
The TVF leverages two Convolutional Neural Networks (CNNs): one to classify the digits, and another one to classify the sizes of the digit. For now, we assume these models are pretrained; in \S\ref{sec:declarative_ml} we describe how we can train them from scratch within the SQL query. 
The output of the TVF are two 2d-tensor columns: one for digit and another one for size. These columns contain the classification probabilities for each tile in the grid encoded using \system's PE API. 
The PE columns are then fed into custom implementations of {\sc group by} and {\sc count} operators 
that are compatible with PE inputs; we describe these operators in more detail in \S\ref{sec:declarative_ml}. 

%
\end{example}

\stitle{\emph{SQL within ML}: Embedding queries in PyTorch programs.}
As described in \S\ref{sec:in-ml-db},  query compilation in \system outputs a PyTorch model. Thus, a compiled query has all the capabilities of PyTorch models. 

\begin{example}[Training Loop]\label{ex:training}
Consider the \mnistgrid query again, except now we want to train the CNNs in the TVF from scratch by providing examples of $\langle \mathit{input},\mathit{output} \rangle$ pairs from our queries. We can simply embed the query within a PyTorch gradient descent training loop, 
as shown in Listing~\ref{lst:training}. 
Note that doing this na\"ively will not work in practice because the SQL query is not end-to-end differentiable. However, in \S\ref{sec:declarative_ml} we will show how we bypass this limitation by introducing \emph{trainable queries}.

\begin{lstlisting}[aboveskip=\smallskipamount,belowskip=\smallskipamount,caption=Training loop for the MNISTGrid query., label={lst:training}]
def train(compiled_query, num_iterations, optimizer, mnist_grids, target_counts): 
    for i in range(num_iterations): 
        optimizer.zero_grad()
        
        # Register MNISTGrid and perform inference with the query
        tdp.sql.register_tensor(mnist_grids[i], "MNIST_Grid")
        predicted_counts = compiled_query.run()
        
        # Compute loss. Here we use MSE between the counts.  
        loss = ((predicted_counts - target_counts[i])**2).mean()
        
        # Backpropagate and perform optimization step
        loss.backward()
        optimizer.step()

optimizer = Adam(compiled_query.parameters(), lr=0.01)
train(compiled_query, 10, optimizer, mnist_grids, target_counts)
        
\end{lstlisting}
\end{example}

\stitle{Declarative, inference-oriented experience.}
We want users to enjoy the full flexibility of PyTorch to express ML transforms, while leveraging SQL to express data operations. In fact, expressing relational operations like {\sc group by} and {\sc count} is unnatural in PyTorch, while implementing the \texttt{parse\_mnist\_grid} TVF in pure SQL is tedious, if not infeasible. The \mnistgrid example demonstrates how we can use the right language for the right task, while seamlessly blending the two in a single unified runtime. We use SQL as a higher-level abstraction or orchestrator between data operations (ingestion, post-processing, relational operators) and machine learning transforms (expressed through UDFs).

This declarative way of expressing hybrid ML-SQL inference can lead to a deployment-first experience, since the query can be directly deployed as-is, i.e., without having to carve out the loss function, training loop, or other components that are only necessary for training, as instead is required in other SQL-first solutions~\cite{DBLP:conf/vldb/Peseux21,10.14778/3317315.3317323}.
Additionally, we believe that it brings improved readability and code sharing across tasks, as well as new perspectives by adding a vocabulary of relational operators in expressing ML inference, as we will show in \S\ref{sec:use-cases}.





\section{Differentiable SQL in \system} 
\label{sec:declarative_ml}

Another powerful implication of using PyTorch as our runtime is that the database has access to automatic differentiation~\cite{pytorch}. \eat{Concretely, automatic differentiation is a tool built within ML frameworks like PyTorch that allows to automatically compute the derivative (or gradients) of a sequence of differentiable operations; 
the tool implements backpropagation~\cite{torch-autograd}, which is the algorithm that enables training neural networks using gradient descent optimization mechanisms. }
Since \system can embed PyTorch ML models within SQL queries using UDFs/TVFs, and has access to automatic differentiation, it introduces a new class of queries, that we refer to as \textit{trainable queries}. This new class of queries: (1)~contains tunable parameters;  
and (2)~can be compiled down to an execution plan that is \textit{end-to-end differentiable}, i.e., a PyTorch model that is composed only of differentiable operators. The latter requirement allows us to backpropagate through the query operators, 
and therefore, train these models using gradient descent optimization schemes. The end result is that, thanks to this feature, users can train any parameters embedded in a trainable query, as we already mentioned in Example~\ref{ex:training}. The feature is enabled in \system by passing a flag at compilation time; Listing~\ref{lst:trainable_query} shows an example using~\mnistgrid.

\lstset{frame=tb,
	language=Python,
	commentstyle=\color{dkgreen},
	aboveskip=3mm,
	belowskip=3mm,
	showstringspaces=false,
	columns=flexible,
	basicstyle={\footnotesize\ttfamily},
	numberstyle=\tiny\color{gray},
	keywordstyle=\color{blue},
	stringstyle=\color{mauve},
	breaklines=true,
	breakatwhitespace=true,
	tabsize=3,
	numbers=none,
	xleftmargin=1em,
	framexleftmargin=1em,
}

\vspace{1ex}
\begin{lstlisting}[aboveskip=\smallskipamount,belowskip=\smallskipamount,caption=Enabling trainable queries in \system., label={lst:trainable_query}]
compiled_query = tdp.spark.query("SELECT Digit, Size, COUNT(*) FROM parse_mnist_grid(MNIST_Grid) GROUP BY Digit, Size", extra_config={tdp.constants.TRAINABLE : True})
\end{lstlisting}
\vspace{0.5ex}

A missing detail here is how to make SQL operators differentiable. For example, the \mnistgrid query contains a {\sc group by} statement with {\sc count} aggregation. It is unusual to think about differentiating traditionally discrete operators like {\sc count}. However, past work has shown that we can often relax discrete operators to continuous, differentiable approximations~\cite{windtunnel, algorithmic_supervision}. A simple example is the logistic function that can approximate a step function while still being differentiable or using \texttt{softmax} as a smooth, differentiable proxy for the \texttt{argmax} function. In our example, we can implement a differentiable count operator on PE data using only addition and multiplication~\cite{rap}. We implement this in PyTorch as our \texttt{soft\_count} operator shown in Fig.~\ref{fig:mnist-grid}, and generalize it to grouped aggregation with our \texttt{soft\_groupby} operator. 
At inference time, we swap the approximate differentiable operators with exact implementations, and thus, eliminate approximation errors.

\vspace{-0.5ex}
\section{Use Cases}
\label{sec:use-cases}
Next, we present some applications and experimental results showcasing \system's integration with PyTorch.
In the experiments we use an Azure NV6 v3 VM equipped with a Intel Xeon CPU E5-2690 v4 @ 2.6GHz (6 virtual cores), and an NVIDIA Tesla V100 GPU. 

\begin{figure*}[t!]
    \centering
    \includegraphics[width=0.88\linewidth]{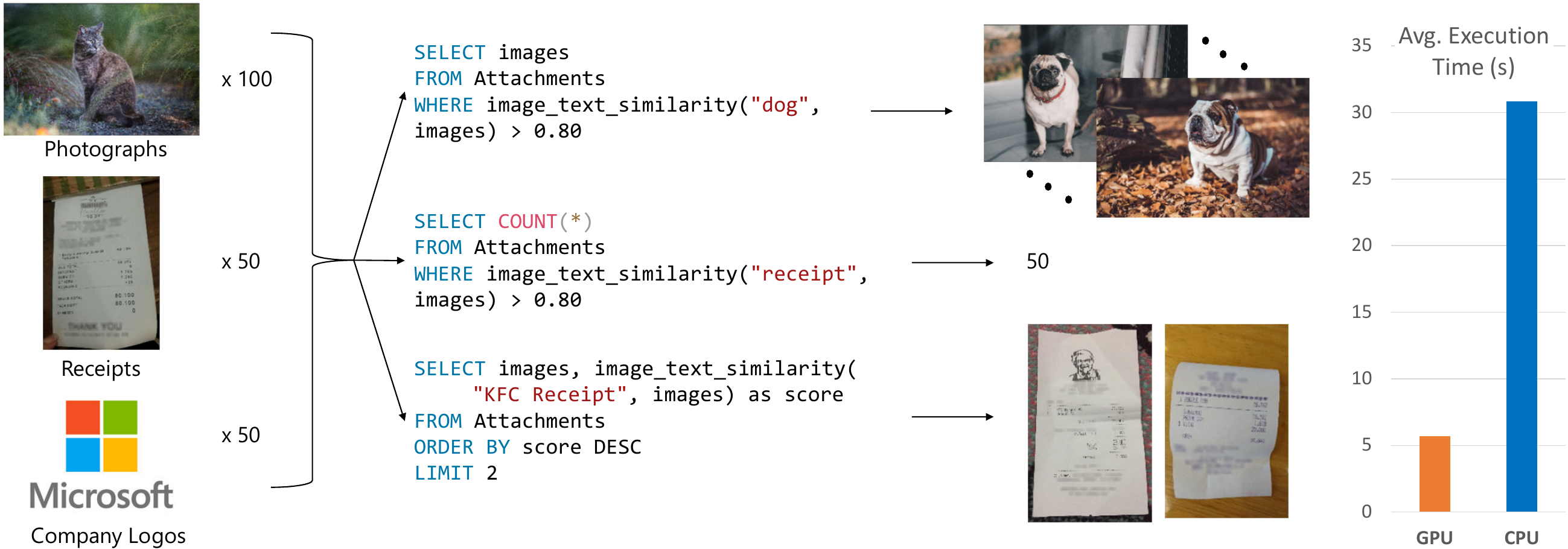}
    \vspace{-2.5ex}
    \caption{(Left) Examples of Multimodal Queries with \system. (Right) Avg. execution time for these queries on 1000 images.}
    \vspace{-1.5ex}
    \label{fig:multimodal}
\end{figure*}



\vspace{-1ex}
\subsection{Multi-modal Queries}

We start by showing an example of how we can use SQL with UDFs to filter or search through images using a natural language criterion. This is similar to vector similarity search~\cite{milvus,analyticdb-v} where similarity between semantic vector representations (or \textit{embeddings}) of queries and search candidates is used to fulfill search queries. However, note that \system provides additional flexibility as, beyond search, we can perform full SQL queries on top of the results of the vector similarity kernel.
To support these multimodal use cases, we create a UDF \texttt{image\_text\_similarity} that computes the similarity score between text and images. This UDF leverages the pre-trained CLIP model~\cite{clip} from OpenAI which is trained to embed images and text with similar semantic meaning to similar vector representations.
Listing~\ref{lst:image_text_similarity} shows how easy it is to leverage and embed state-of-the-art pre-trained models within a query through  \system 's UDFs. 

\begin{lstlisting}[caption=UDF to compute similarity scores between a natural language query and a column of images., label={lst:image_text_similarity}]
from transformers import CLIPProcessor, CLIPModel
model = CLIPModel.from_pretrained("openai/clip-vit-base-patch32")
processor = CLIPProcessor.from_pretrained("openai/clip-vit-base-patch32")

@tqp_udf("float")
def image_text_similarity(query: str, images: torch.Tensor) -> torch.Tensor: 
    inputs = processor([query], images, return_tensors="pt", padding=True)
    outputs = model(**inputs)
    scores = outputs.logits_per_image.flatten() / 30
    return scores
\end{lstlisting}

As an example application, suppose we are trying to run queries against a dataset of email image attachments. Fig.~\ref{fig:multimodal} shows a sample dataset of email attachments created from 100 images of photographs, 
50 receipts, 
and 50 company logos. 
In the middle, we see three examples of multimodal queries we may want to run on this dataset. From top to bottom: the first query implements a filter query on the attachments; the second combines relational aggregate operations on top of the results of the filter query; finally, the third query implements a top-k image search query as is common in vector similarity search engines like Milvus \cite{milvus}.

Since \system can seamlessly leverage PyTorch for GPU acceleration, we compare the performance on CPU and GPU. Specifically, we run a workload of 30 queries containing a mix of queries as shown in Fig.~\ref{fig:multimodal} on a dataset of 1,000 200x300 images, and measure the average query execution time. Fig.~\ref{fig:multimodal} (right) shows the results with GPU execution being around 5$\times$ faster. We are currently integrating approximate indexing~\cite{milvus} into \system for speeding up top-k queries.



\vspace{-1.2ex}
\subsection{SQL Queries over OCRed Documents}



\begin{figure*}[t!]
\centering
\includegraphics[width=0.95\linewidth]{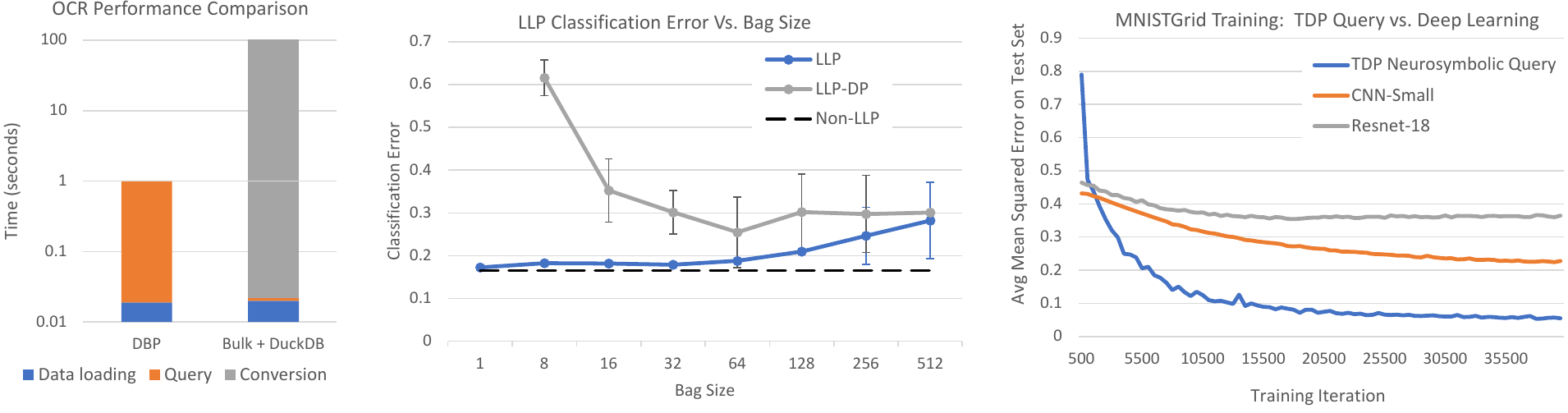}\vspace{-2.5ex}
\caption{Experiment Results: OCR (Left, Section 5.2), LLP (Middle, Section 5.3), MNISTGrid (Right, Section 5.4).}
\label{fig:results}\vspace{-1.5ex}
\end{figure*}

Next, we push the boundary a little bit further and show how with \system we can execute SQL queries over tables extracted from images. 
Specifically, in this scenario we start with a set of images and related metadata, and we want to execute queries over the data stored into the tables in the images, and filter the images based on some metadata information.
This scenario non-trivially mixes scalar filters with operations over multidimensional data storing the raw images.
We implemented this scenario by generating 100 images, using the {\sf dataframe\_image} Python library, from Pandas dataframes of the Iris dataset, 
and attaching to each a timestamp specifying when the image was generated.
We then load the data in \system and query them as shown in listing~\ref{lst:ocr}.

\vspace{-0.3ex}
\begin{lstlisting}[aboveskip=\smallskipamount,belowskip=\smallskipamount,caption=Querying tables stored on {\sf Document} images., label={lst:ocr}]
SELECT AVG(SepalLength), AVG(PetalLength)
FROM (SELECT extract_table(images)
    FROM Document WHERE timestamp = "2022:08:10")
\end{lstlisting}
\vspace{0.2ex}

This query fetches a single image using the filter over the timestamp, and computes the average over two columns. 
To extract the data from the tables we use a UDF, \texttt{extract\_table}, which internally employs a pipeline of ML models to: (1)~recognize where the table is in the image; and (2)~OCR the image and convert it into a plain tensor. Note that the query above is executed end-to-end on GPU in \system.
As far as we know, no other database system is able to support such scenario natively, so we compared our implementation to a version in which all images are first run through the models for extracting the data, and then the data is loaded into a DuckDB instance and queried.
As we can see from Fig.~\ref{fig:results} (left), our approach is 2~orders of magnitude faster, because we only require to convert a single image, instead of bulk-convert all of them. Note that loading the raw images in \system takes approximately the same time as saving and loading the extracted table data into DuckDB. Conversely, DuckDB query execution time is only few milliseconds, while in our case it requires around 1~second to fetch the image, convert it, and query it. Data conversion takes the majority of the time.

\subsection{Learning from Label Proportions (LLP)}

Learning from Label Proportions (LLP)~\cite{onLLP} is an ML problem setting where we learn from proportions (or equivalently counts) of classification labels over a set of instances. More specifically, in LLP, training data comes in the form of \textit{bags}, where each bag is a collection of instances. 
The goal is to train a classifier on the individual instances in the bag, given only the aggregated count annotations per bag. LLP has a broad set of real-world applications, including: learning from medical data where often the standard practice is to release counts instead of individuals' data for privacy \cite{llp_embryo, llp_medical_data}, learning in settings with instrumentation limitations such as high-energy physics where aggregates observations may be more reliable than instance-level observations \cite{llp_physics, llp_spms}, 
learning from noisy counts for label differential privacy \cite{llp_dp},
learning from aggregates where instance-level labels are much more expensive to collect \cite{llp_image, llp_video}, and learning from aggregate clickstream data \cite{llp_clicks}. 

%

Interestingly, 
SQL provides natural declarative syntax to model the process of obtaining count labels for bags in LLP. A user can simply provide a UDF to classify instances in a bag and then use a {\sc group-by-count} query to obtain the counts for each class in the bag.
 As an example, we show an application of SQL for LLP using the Adult Income Dataset,
 as has been explored in past LLP literature \cite{onLLP}. 
 The Adult Income Dataset includes a subset of the 1994 US Census data, where 
 for each record, there is an associated binary classification label indicating whether the individual's income is >50K. 
 While the dataset does provide labels per individual it is common to use this dataset to benchmark LLP methods by generating bags with classification label counts at different granularities. 
 When training, we provide only aggregate count labels per bag and not the individual labels. We perform experiments by varying the bag sizes $\in \{1, 8, 16, 32, 128, 256, 512\}$. For testing, we compute the classification error on the individual labels in the test set. 
 The SQL query and the TVF are presented in Listing~\ref{lst:adult}.

\begin{lstlisting}[aboveskip=\smallskipamount,belowskip=\smallskipamount,caption=Linear classifier TVF and \system query used to implement LLP inference on the Adult Income Dataset., label={lst:adult}]
linear_model = torch.nn.Linear(len(num_feature_cols, 2))

@tdp_udf("Income float")
def classify_incomes(x: torch.Tensor) -> torch.Tensor:
    return linear_model(x)
    
query = tdp.spark.query("SELECT Income, COUNT(*) FROM classify_incomes(Adult_Income_Bag) GROUP BY Income", extra_config={tqp.constants.TRAINABLE : True})
\end{lstlisting}


The blue LLP line in Fig.~\ref{fig:results} (middle) shows the classification errors of our experiments. For comparison, the flat dashed (Non-LLP) line shows the results of training a model in a typical classification setting, where individual classification labels are available for training. Observe that the LLP experiment errors are quite close to the Non-LLP results for small bag sizes. As is typical in LLP~\cite{onLLP}, the error gradually increases as we increase the bag size, since with larger bag sizes, we dilute the finer, instance-level signal. Still, the error remains relatively stable even for relatively large bag sizes. 

\subsection{Label Differential Privacy with LLP}
Learning from aggregates lends itself well to privacy-preserving ML. The gold standard of privacy today is differential privacy (DP), a mathematical framework that defines and provides privacy guarantees for algorithms that access data \cite{dp_dwork2006, dp_dwork2014}. 
A commonly used mechanism in differential privacy is to add noise to a query's answer. 
%
When differential privacy is applied to machine learning, privacy of both the features and the labels must be preserved. This kind of all-or-nothing privacy, while powerful, may not be strictly necessary in certain settings. Furthermore, existing learning methods that comply with differential privacy often suffer considerably in model accuracy compared to their non-private counterparts \cite{label_dp}. For these reasons, there has been a search for alternative definitions of privacy. 

One such standard is \textit{label differential privacy} (or Label-DP), which relaxes differential privacy to only apply to the labels in a dataset \cite{label_dp}: there are settings where we care about the privacy of labels but not necessarily the features. For example, in a university student survey asking about vaccination status, while the student information may already be publicly available, the vaccination status is sensitive and should remain private. Similarly in the Adult Income census data described above, we may deem the income level classification labels as sensitive, while treating the other features as not sensitive. Previous work has extended LLP with the Laplace mechanism to learning from noisy counts in order to preserve label differential privacy \cite{llp_dp}. We apply the same approach to the Adult Income Dataset, learning from noisy counts with our trainable SQL query, instead of actual counts. Following \cite{llp_dp}, we set the privacy loss parameter $\epsilon = 0.1$ where $\epsilon$ controls the scale of the Laplace noise added to the count labels.

The gray LLP-DP line in Fig.~\ref{fig:results} (middle) shows the results of our experiments. Here, for small bag sizes, the error is very high, as the noise overpowers the label signal. This is expected as a smaller bag size, requires a higher proportion of noise to ensure to the privacy of individuals. For larger bag sizes, just as in the non-noisy LLP case, we see a gradual increase in error due to the dilution of individual label information through aggregation. Thus, for LLP-DP there is a trade-off between these two factors, with optimal bag size in our case being 64. 


\vspace{-1ex}
\subsection{Learning to Answer Queries over Images}
\label{sec:query-on-images}

We revisit the \mnistgrid example, that we introduced in \S\ref{sec:udf} and \S\ref{sec:declarative_ml}, to demonstrate the combination of unstructured data processing and differentiable SQL capabilities of \system. 
This example has connections to LLP since we are supervising from label counts too. However, \mnistgrid generalizes the LLP Adult Income example in three ways: (1)~we perform the query on images rather than tables; (2)~we group the counts by more than one class; (3)~we train a multi-class classifier for each class rather than just a binary classifier. 
This generalizability is one of the strengths of the SQL abstraction.
In addition, our approach to solve the \mnistgrid example also has relations to neurosymbolic programming (e.g.,~\cite{neurosymbolic_vqa, scallop}). We use our TVF (with \textit{neural} networks) to parse the image into a structured (or \textit{symbolic}) representation. This symbolic representation is then further processed by relational operators. In this way, we can think of \system's SQL as being a declarative language for expressing neurosymbolic computation that 
can be made end-to-end differentiable.

Past neurosymbolic works \cite{neurosymbolic_vqa, scallop} 
have found that by embedding symbolic knowledge into the model inference leads to better training efficiency and generalization. To understand this better, let us consider the alternative approach: modeling this problem as a multiple regression problem using deep learning. Here we treat the 20 grouped counts as regression outputs, and train one CNN to predict these outputs. There are a few disadvantages to this approach compared to the neurosymbolic approach: (1) the CNN must learn not only how to classify the tiles but also learn the {\sc group by} and {\sc count} operations from scratch making training less efficient; (2) since this uses a single, monolithic CNN to learn the whole query task, it entangles learning of the classification and relational operations, disallowing generalization to other tasks (something we purposely try to avoid in \system, as previously described in \S\ref{sec:udf}). 
Next we describe two experiments showing the above differences. \vspace{1ex} 

\stitle{Experiment 1: More efficient training.}
We compare the training behavior of our approach against two pure deep learning models: (1) CNN-Small with 850K trainable parameters; and (2) Resnet-18  
with 11.1M trainable parameters. We choose the first as it has similar architecture to the CNNs we use in the \mnistgrid TVF, and has similar number of trainable parameters. 
We choose Resnet-18 as it is often used as the backbone architecture in state-of-the-art CNNs. We train each of the three approaches for 40,000 iterations and with similar hyperparameters, and our results are the average of 5 runs. The training set contains 5,000 images while the test set contains 1,000 images. Fig~\ref{fig:results} (right) plots the \mnistgrid test error for each training iteration for the three approaches. As expected, the \system neurosymbolic approach converges to a close-to-zero error very quickly. The two deep learning approaches learn much slower and asymptote to much higher errors compared to the \system approach. \vspace{1ex}

\stitle{Experiment 2: Better generalization.}
As shown in Fig~\ref{fig:mnist-grid}, our approach allows to decompose the query execution into clear subcomponents, any one of which can be reused in a future query without losing generality. For example, after training the query on the \mnistgrid task, we can pull out the trained \texttt{digit\_parser} CNN and embed it into a completely new query that requires digit classification. 
To show this, we extract the \texttt{digit\_parser} CNN from our \system query and test the classification performance of this model on the MNIST dataset: the model achieves 98.15\% classification accuracy on average, without ever explicitly being trained using the MNIST classification labels.

\section{Related Work}


Integrating ML and databases is a research area that has received a lot of attention for quite some time. While early works tried to optimize the hand-off of data between separate ML and DB systems~\cite{madlib,10.14778/3467861.3467867}, recently we are seeing more works trying to execute ML workloads directly on databases, e.g., decomposing ML operations into SQL queries~\cite{10.1145/3468791.3468840,masq,DBLP:conf/vldb/Peseux21,10.14778/3317315.3317323}. Differently than previous approaches, we instead propose a data platform built on TCRs. Therefore, by construction, we are able to leverage features such as hardware acceleration, differentiability, and multi-modal data support, as well as native training and prediction of ML models.
While similarly to previous approaches we leverage UDFs/TVFs to switch to ML operations from within SQL, our context switch does not introduce any overhead since functions and SQL operators are executed on the same tensor runtime. Furthermore, we find this approach to be more usable for data scientists rather than a SQL-centric approach where both ML and relational operations are expressed directly in SQL.

In the past few years, we have been working on several projects targeting mixed SQL/ML workloads~\cite{raven-sigmod,raven,tqp,tqp-demo}, and how ML systems can be leveraged beyond pure deep learning workloads~\cite{hummingbird,hummingbird-vision}. \system is the next step in this journey. 
We believe that AI-centric database systems are required to target the next generation of data-driven applications such as agriculture~\cite{9705193}, chemistry~\cite{williams2021evolution}, and photo fraud detection~\cite{milvus-photo-fraud}.
Custom-built data management systems (e.g.,~\cite{milvus,viva,analyticdb-v}) have been proposed lately to address this new set of challenges. 
Conversely, we believe that AI-centric database systems can be flexible enough to support these new workloads while also being performant on more legacy ones~\cite{tqp}.

\section{Conclusion}

There is a lot of ongoing excitement around Data-centric AI~\cite{data-centric-ai}. In contrast, in this paper, we are proposing an AI-centric database.
In fact, we demonstrate that a tight integration between ML and SQL can bring value to both the ML and the database communities.
For the database community, building a database on a tensor runtime allows to leverage hardware acceleration, differentiability and multi-modal data.
For the ML community, we showed how complex tasks such as LLP can be declaratively expressed in SQL.
While we are in the early stages of this journey, we are excited about the potential benefits that an AI-centric database system can bring to end users.

\end{sloppypar}

\balance
\vspace{-0ex}
\bibliographystyle{ACM-Reference-Format}
\bibliography{paper}

\end{document}